\documentclass[letter]{aa} 
\usepackage{graphicx}
\usepackage{amsmath}    
\usepackage{amssymb}    
\usepackage{txfonts}
\usepackage{natbib,twoopt}
\usepackage[breaklinks=true]{hyperref} 

\hypersetup{
  colorlinks=true,   
  citecolor=blue,    
  urlcolor=red,     
  linkcolor=blue,     
}

\bibpunct{(}{)}{;}{a}{}{,} 

\newcommand{\EL}[1]{{ #1}}

\begin{document} 

   \title{Tension with the flat $\Lambda$CDM model from a high-redshift Hubble diagram of supernovae, quasars, and gamma-ray bursts}

   \author{E. Lusso\inst{1,2}\thanks{\email{elisabeta.lusso@unifi.it}}, E. Piedipalumbo\inst{3,4}, G. Risaliti\inst{1,2}, M. Paolillo\inst{3,4,5}, S. Bisogni\inst{2}, E. Nardini\inst{2} \and L. Amati\inst{5}. 
          }
          
\institute{
$^{1}$Dipartimento di Fisica e Astronomia, Universit\`a di Firenze, via G. Sansone 1, 50019 Sesto Fiorentino, Firenze, Italy\\
$^{2}$INAF--Osservatorio Astrofisico di Arcetri, 50125 Florence, Italy\\
$^{3}$Dipartimento di Fisica, Universit\`a degli studi di Napoli Federico II, via Cinthia, 80126 Napoli, Italy\\
$^{4}$INFN - Sezione di Napoli, via Cinthia 9, 80126 Napoli, Italy\\
$^{5}$INAF - Osservatorio Astronomico di Capodimonte, via Moiariello 16, 80131 Napoli, Italy\\
$^{6}$INAF - Osservatorio di Astrofisica e Scienza dello Spazio, via P. Gobetti 93/3, Bologna, Italy
}

  \titlerunning{Tension on the $\Lambda$CDM model using SNe, QSOs and GRBs}
  \authorrunning{Lusso E. et al.}
   \date{Received July 3, 2019; accepted July 17, 2019}

 
  \abstract{In the current framework, the standard parametrization of our Universe is the so-called Lambda Cold Dark Matter ($\Lambda$CDM) model. Recently, a $\sim$4\,$\sigma$ tension with the $\Lambda$CDM model was shown to exist via a model-independent parametrization of a Hubble diagram of type Ia  supernovae  (SNe Ia) from the JLA survey and quasars. 
Model-independent approaches and independent samples over a wide redshift range are key to testing this tension and any possible systematic errors. Here we present an analysis of a combined Hubble diagram of SNe Ia, quasars, and gamma-ray bursts (GRBs) to check the agreement of the quasar and GRB cosmological parameters at high redshifts ($z>2$) and to test the {\it concordance} flat $\Lambda$CDM model with improved statistical accuracy. We build a Hubble diagram with SNe Ia, quasars , and GRBs, where quasars are standardised through the observed non-linear relation between their ultraviolet and X-ray emission and GRBs through the correlation between the spectral peak energy and the isotropic-equivalent radiated energy (the so-called {\it Amati relation}). We fit the data with cosmographic models consisting of a fourth-order logarithmic polynomial and a fifth-order linear polynomial, and compare the results with the expectations from a flat $\Lambda$CDM model. We confirm the tension between the best-fit cosmographic parameters and the $\Lambda$CDM model at $\sim$$4\sigma$ with SNe Ia and quasars, at $\sim$$2\sigma$ with SNe Ia and GRBs, and at $>4\sigma$ with the whole SNe Ia+quasars+GRB data set. The completely independent high-redshift Hubble diagrams of quasars and GRBs are fully consistent with each other, strongly suggesting that the deviation from the standard model is not due to unknown systematic effects but to new physics.}

   \keywords{quasar: general -- supernovae: general -- gamma-ray burst: general -- cosmology: cosmological parameters, observational}

   \maketitle
%

\section{Introduction}
Diverse sets of cosmological probes have now established that our Universe is undergoing an accelerated expansion: from local type Ia supernovae (SNe Ia; e.g. \citealt{riess1998,perlmutter1999,suzuki2012}) to the combined constraints of the cosmic microwave background (CMB), baryonic acoustic oscillations (BAO), SNe \EL{Ia}, and lensing provided by the \citet{planck2018}. The observed accelerated expansion is ascribed to the so-called {\it dark energy} ($\Omega_\Lambda$), which is found to be $\sim70\%$ (within the standard {\it concordance model}) of the energy budget, thus the dominant component of today's Universe. 
In the basic concordance model (i.e. $\Lambda$CDM), acceleration is driven by a cosmological constant, $\Lambda$, with an additional cold dark matter ($\Omega_M$) component in a flat Universe. Although this model has been overall successful in fitting the data, observational evidence suggesting a deep revision of its underlying assumptions is now mounting. 
For example, the determination of the Hubble constant ($H_0$), which represents the normalisation of the Hubble parameter $H(z)$, from local sources (Cepheids, supernovae Ia) is in tension at 4.4$\sigma$ with respect to the one from the sound horizon observed from the CMB 
 \citep{riess2019}. If systematic uncertainties are not the main driver of this discrepancy, a significant disagreement would call for fundamental physics beyond the standard model (e.g. time-dependent dark energy equation of state, modified gravity, additional relativistic particles).

Recently, model-independent measurements of the distance modulus--redshift relation (the so-called {\it Hubble diagram}) of quasars combined with SNe Ia from the \EL{\textit{Sloan Digital Sky Survey}-II/\textit{Supernova Legacy Survey} 3} (SDSS-II/SNLS3) {\it Joint Light-Curve Analysis} (JLA, \citealt{betoule2014}) have shown a deviation from the $\Lambda$CDM model emerging at high redshifts ($z>1.4$) with a statistical significance of $\sim4\sigma$ \citep[RL19 hereafter]{rl19}. 

In the present work, we present improved measurements of the Hubble diagram leveraging on the analysis presented by RL19. By combining the quasar sample of these latter authors with the most updated compilation of SNe Ia from the {\it Pantheon} survey \citep{scolnic2018} and publicly available gamma-ray burst data \citep{demianski17a}, we are now finding a tension between the best-fit cosmographic parameters and the $\Lambda$CDM model at $>4\sigma$. 

\section{The data set}
\begin{figure}
 \centering
  \includegraphics[width=\linewidth,clip]{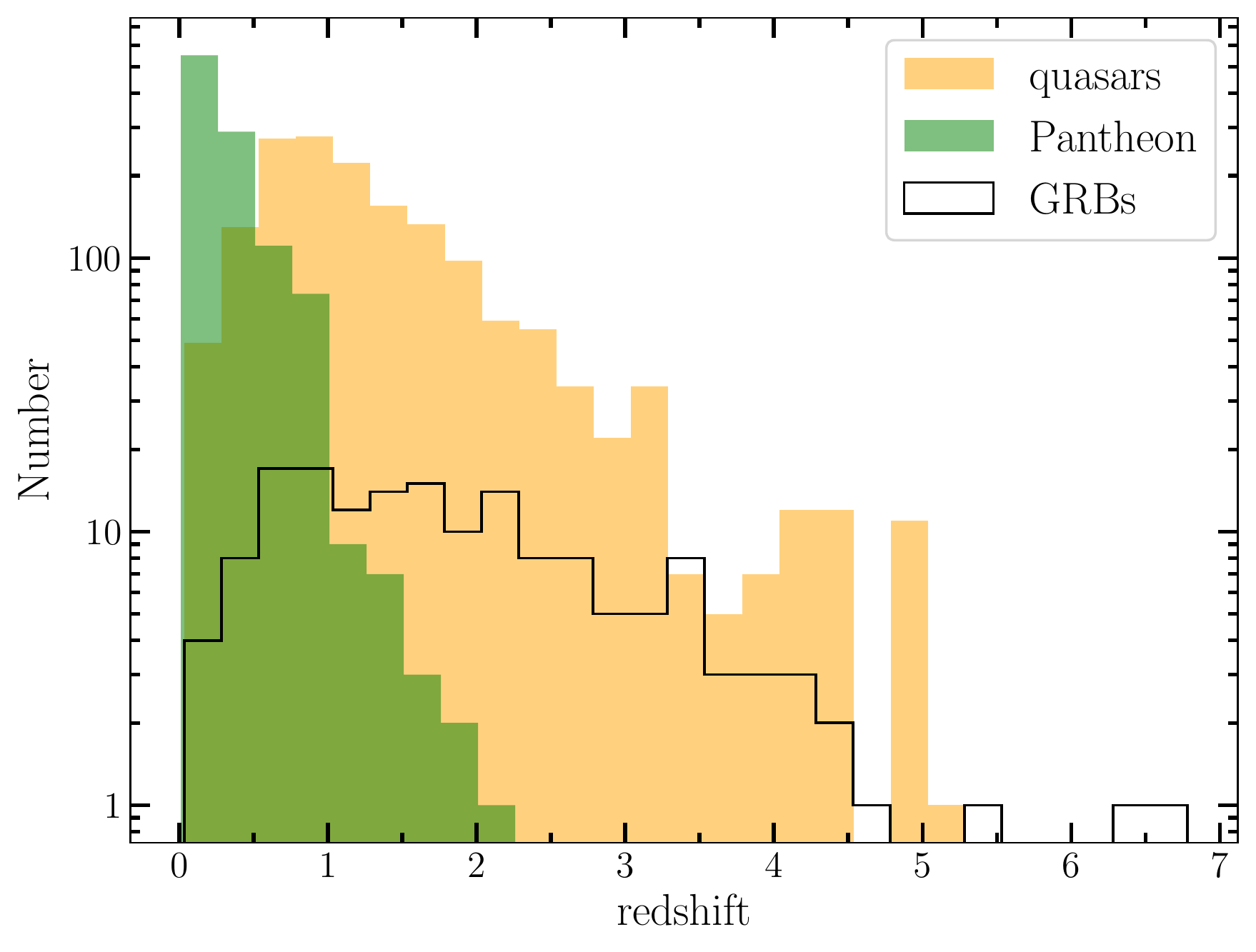}
        \caption{Redshift distribution of the {\it Pantheon} SNe Ia sample (green, \citealt{scolnic2018}), the quasars sample (yellow, \citealt{rl19}), and the GRB sample (open black, \citealt{demianski17a}).}
  \label{zdistrib}
\end{figure}

To compute the cosmological parameters, we built a Hubble diagram by combining three samples:
\begin{enumerate}
\item {\it Pantheon}. A SNe Ia sample consisting of 1048 objects in the range $0.01 < z < 2.26$, significantly extending the redshift coverage at $z>1$ with respect to JLA (i.e. maximum redshift $z\simeq1.3$). This sample is a combination of 365 spectroscopically confirmed SNe Ia discovered by the Pan-STARRS1 (PS1) Medium Deep Survey together with the subset of 279 PS1 SN Ia ($0.03 < z < 0.68$) with distance estimates from SDSS, SNLS, and various low-redshift and HST samples (see Table 4 in \citealt{scolnic2018}). For further details on photometry, astrometry, calibration, and systematic uncertainties on SN Ia distances we refer the interested reader to \citet{scolnic2018}.

\item {\it Quasars.} The quasar sample is composed of 1598 sources in the redshift range $0.04<z<5.1$ with high-quality UV and X-ray flux measurements. Distance moduli are computed from the non-linear relation between the ultraviolet and the X-ray emission observed in quasars \citep{2015ApJ...815...33R,2016ApJ...819..154L}. All the details on sample selection, X-ray, and UV flux computation and the analysis of the non-linear relation, calibration, and a discussion on systematic errors are provided in RL19.

\item {\it Gamma ray bursts}. This sample is a compilation of 162 GRBs ranging from $0.03 < z < 6.67$. Due to convergency issues of the standard cosmographic approach \citep{capozziello2011}, we limited the GRB redshift range to $z=6.67$, thus excluding the two highest redshift objects at $z=8.1$ and $z=9.3$.
The Hubble diagram is constructed by calibrating the correlation between the peak photon energy, $E_{\rm p,i}$, and the isotropic equivalent radiated energy, $E_{\rm iso}$ \citep{amati2008}. We refer to \citet{demianski17a,demianski17b} and \citet{amatidellavalle13} for details on the data set and on the calibration of the distance modulus--redshift relation, as well as for discussion and analysis of possible instrumental and selection effects (see also \citealt{amati2006,ghirlanda2008,nava2012,amati2019}).

\end{enumerate}
Figure~\ref{zdistrib} shows the redshift distribution of the three samples used to build the Hubble diagram. 
The combination of GRBs and quasars is such that we can probe a redshift range better suited for studying dark energy than the one covered by SNe Ia. Nevertheless, SNe Ia are necessary to cross-calibrate both quasars and GRBs to compute absolute distances.

\section{Fitting the hubble diagram: a cosmographic approach}
In this analysis, our goal is to use the Hubble diagram obtained by combining SNe Ia, quasars, and GRBs to test the {\it concordance} flat $\Lambda$CDM model through a model-independent cosmographic method. The cosmographic approach to cosmology is receiving increasing interest as it can be used to obtain a direct fit of the Hubble diagram without any hypothesis concerning the type of dark energy and dark matter required to satisfy the Einstein equation, and simply assuming the minimal priors of isotropy and homogeneity. Here we describe two cosmographic techniques we applied to the data in order to measure the cosmological parameters and test the concordance model.

\subsection{Traditional cosmographic approach}
\label{tradcosmo}
To compute cosmological parameters, we followed the standard cosmographic procedure which is fully described in Sect. 2 of \citet{demianski17b}. We briefly summarise the main steps.
We parametrized the space-time geometry described by the Friedmann-Robertson-Lema{\^i}tre-Walker (FLRW) metric by a Taylor series of the scale factor $a(t)$ and its higher-order derivatives. 
This expansion leads to a distance--redshift relation that only relies on the assumption of the FLRW metric and is therefore fully model independent. From the cosmographic functions (\citealt{visser2004} and equations from 2 to 6 in \citealt{demianski17b}), we can compute the series expansion to the fifth order in time of the scale factor
\begin{eqnarray}\label{eq:a_series}
\frac{a(t)}{a(t_{0})} &=& 1 + H_{0} (t-t_{0}) -\frac{q_{0}}{2}
H_{0}^{2} (t-t_{0})^{2} +\frac{j_{0}}{3!} H_{0}^{3} (t-t_{0})^{3}
\\ \nonumber &+&\frac{s_{0}}{4!} H_{0}^{4} (t-t_{0})^{4}+ \frac{l_{0}}{5!}
H_{0}^{5} (t-t_{0})^{5} +\emph{O}[(t-t_{0})^{6}]\,,
\end{eqnarray}
where $q_0$, $j_0$, $s_0$, and $l_0$ are commonly indicated as the deceleration, jerk, snap, and lerk parameters, corresponding to the functions evaluated at the present time $t_0$ \citep{sahni2003}.
With this metric, it is possible to express all the observational quantities, like the luminosity
distance $d_L$,  as a power series in the redshift parameter $z$.
From Eq. (\ref{eq:a_series}), we can construct the series for the luminosity distance as
\begin{eqnarray}\label{serielum1}
D_{L}(z) = \frac{c z}{H_{0}} \left( \mathcal{D}_{L}^{0} +
\mathcal{D}_{L}^{1} \ z + \mathcal{D}_{L}^{2} \ z^{2} +
\mathcal{D}_{L}^{3} \ z^{3} + \mathcal{D}_{L}^{4} \ z^{4} +
\emph{O}(z^{5}) \right)\,,
\end{eqnarray}
where {\setlength\arraycolsep{0.2pt}
\begin{eqnarray}\label{serieslum2}
\mathcal{D}_{L}^{0} &=& 1\,, \\
\mathcal{D}_{L}^{1} &=& - \frac{1}{2} \left(-1 + q_{0}\right)\,, \\
\mathcal{D}_{L}^{2} &=& - \frac{1}{6} \left(1 - q_{0} - 3q_{0}^{2} + j_{0}\right)\,, \\
\mathcal{D}_{L}^{3} &=& \frac{1}{24} \left(2 - 2 q_{0} - 15
q_{0}^{2} - 15 q_{0}^{3} + 5 j_{0} + 10 q_{0} j_{0} + s_{0} \right)\,,\\
\mathcal{D}_{L}^{4} &=& \frac{1}{120} \left( -6 + 6 q_{0} + 81
q_{0}^{2} + 165 q_{0}^{3}  -105 q_{0}^{4} - 110 q_{0} j_{0}  +\right.\nonumber \\ \nonumber - && \left.105
q_{0}^{2} j_{0} - 15 q_{0} s_{0} - 27 j_{0} + 10 j_0^{2} - 11 s_{0} - l_{0}\right)\,.
\end{eqnarray}}
To mitigate the usual convergence problems of the cosmographic series expansion at high redshifts, we consider the expansion in the improved parameter $y = z/(1+z)$ (the so-called $y-$parameter), which has the great advantage of also holding for $z > 1$ and  is therefore the appropriate tool for handling high-redshift sources such quasars and GRBs \citep{catton2007,vitagliano2010, demianski12,demianski17b}.
 
\subsection{Cosmographic parameters from a logarithmic polynomial function.}
\label{ourcosmo_theory}
Following a procedure first described in RL19, we fitted the data with a fourth-grade polynomial of $\log(1+z)$ and compare the best fit parameters with the expectations from the standard model. 
In particular, we define a cosmographic function of the luminosity distance as:
\begin{equation}
  \begin{gathered}
  \label{dl}
D_L(z)=k \ln(10)\frac{c}{H_0}\times[\log(1+z)+a_2\log^2(1+z)+\\
        +a_3\log^3(1+z)+a_4\log^4(1+z)] +\emph{O}[\log^5(1+z)],
  \end{gathered}
\end{equation}
where $k, a_2, a_3$, and $a_4$ are free parameters. The parameter $k$ is fitted separately for SNe Ia, quasars, and GRBs in order to cross-calibrate the samples. We note that the Hubble constant $H_0$ is degenerate with the $k$ parameter, so we can fix it to an arbitrary value ($H_0=70$~km s$^{-1}$Mpc$^{-1}$).
The best-fit parameters and their uncertainties were obtained through the Python package {\it emcee} \citep{2013PASP..125..306F}, which is a pure-Python implementation of Goodman \& Weare's affine invariant Markov chain Monte Carlo (MCMC) ensemble sampler. 

Within the standard model, the only free parameter beyond $H_0$ is the matter density, $\Omega_M$. Therefore, we can expand the function $D_L(\Omega_M,z)$ in a series of $\log(1+z)$, and obtain the coefficients of each term as a function of $\Omega_M$. This procedure is discussed in detail by RL19 for a third-order expansion, which is shown to reproduce a flat $\Lambda$CDM model up to $z>5$, with only a few per cent difference at the upper end of the redshift range (see their Supplementary Figure~5). Since the inclusion of GRBs requires a precise fit up to $z\sim6.7$, here we extend the expansion to the fourth order. In this way, our logarithmic expansion can match the standard model in the whole redshift range with a maximum difference below 2\%. After simple algebra we obtain
\begin{eqnarray}
  \begin{gathered}
\label{cosmoparam}
        a_2= \ln(10)\times\left(\frac{3}{2}-\frac{3}{4}\Omega_M\right)  \\
        a_3= \ln^2(10)\times\left(\frac{9}{8}\Omega_M^2-2\Omega_M+\frac{7}{6}\right)  \\
        a_4= \ln^3(10)\times\left(-\frac{135}{64}\Omega_M^3+\frac{18}{4}\Omega_M^2-\frac{47}{16}\Omega_M+\frac{5}{8}\right).
  \end{gathered}
\end{eqnarray}

The above equations describe a curve in the ($a_2, a_3, a_4$) parameter space that we can directly compare with the results of the MCMC technique applied to the SNe+quasars+GRBs Hubble diagram.

\begin{table}
\begin{center}
 \centering
  \setlength{\tabcolsep}{0.1 em}
\begin{tabular}{ccccc}
  \hline
\hline
  Parameter&$q_0$&$j_0$&$s_0$\\
  \hline
  Best fit &$-0.802$ & $3.02$ & $5.9$ \\
  Mean & $-0.8$&$ 3.04$&$5.6$ \\
  2 $\sigma$ & $ (-0.95, -0.66)$&$ (2.15, 4.4)$&$(3.7,9.8)$\\
  \hline
\end{tabular}
\end{center}
\caption{Constraints on the cosmographic parameters (see Sect. \ref{stdapproach}) from
the joint analysis of the SNe Ia, GRB, and QSO Hubble diagram.}
\label{tabgrb}
\end{table}
\begin{figure}
 \centering
  \includegraphics[width=\linewidth,clip]{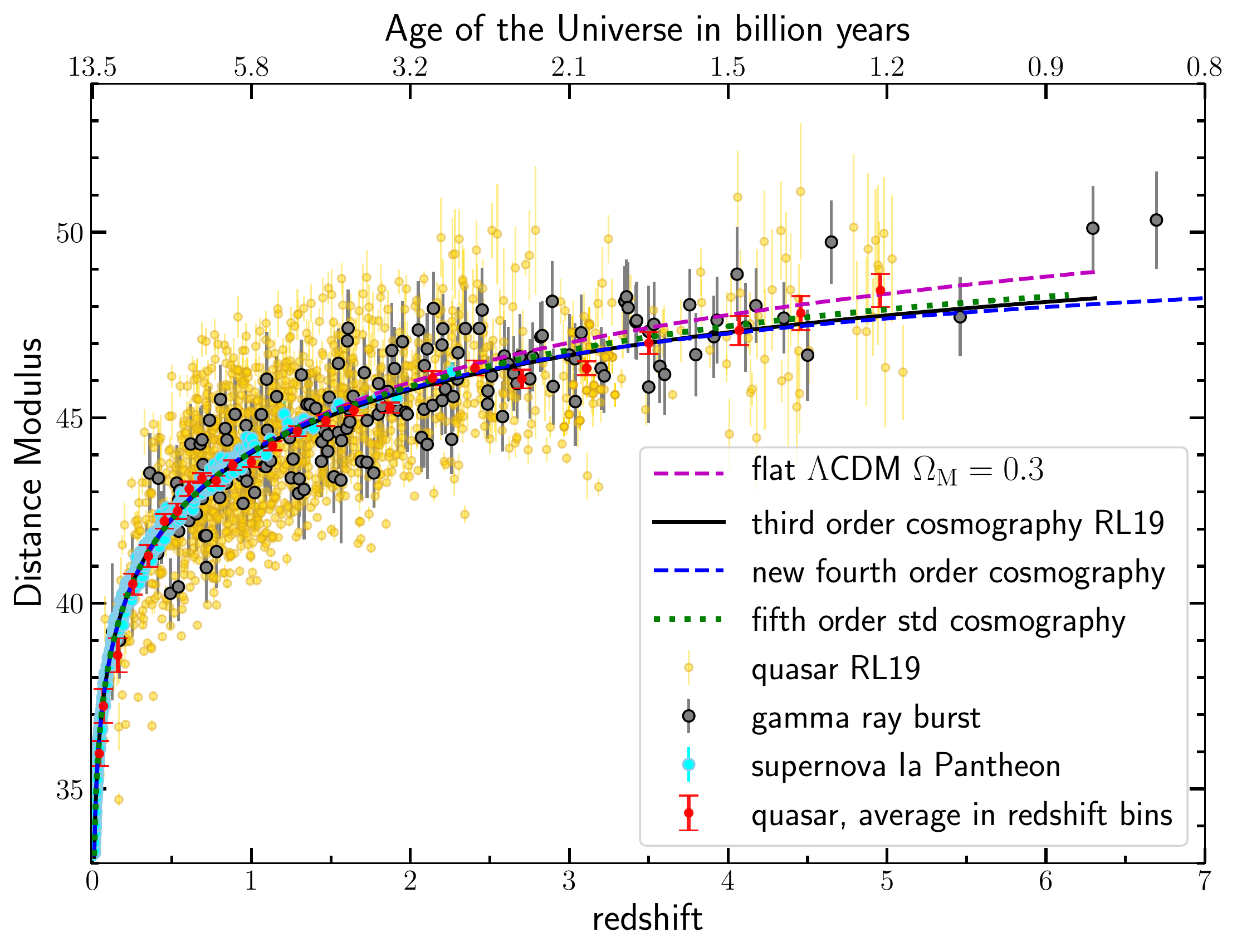}
        \caption{Hubble diagram of SNe Ia (cyan points), quasars (yellow points), and GRBs (black points) with the 1$\sigma$ uncertainties.
Red points are the mean (also with 1$\sigma$ uncertainties) of the distance modulus in narrow redshift bins for quasars only (shown for visualisation purposes). 
The dashed magenta line shows a flat $\Lambda$CDM model fit with $\Omega_{\rm M}= 0.3$. The green dotted line is the best MCMC fit using the standard cosmographic approach (Sect.\,\ref{tradcosmo}) with the inclusion of GRBs and {\it Pantheon}, whilst the blue dot-dashed line represents the new regression fit (Sect.\,\ref{ourcosmo_theory}). The black solid line is the best MCMC regression curve of the third-order expansion of $\log(1+z)$ published by RL19 and obtained by fitting the quasars and the SNe Ia from the JLA survey \citep{betoule2014}.}
  \label{hubblediag}
\end{figure}

\begin{figure}[t!]
 \centering
 \includegraphics[width=0.75\linewidth]{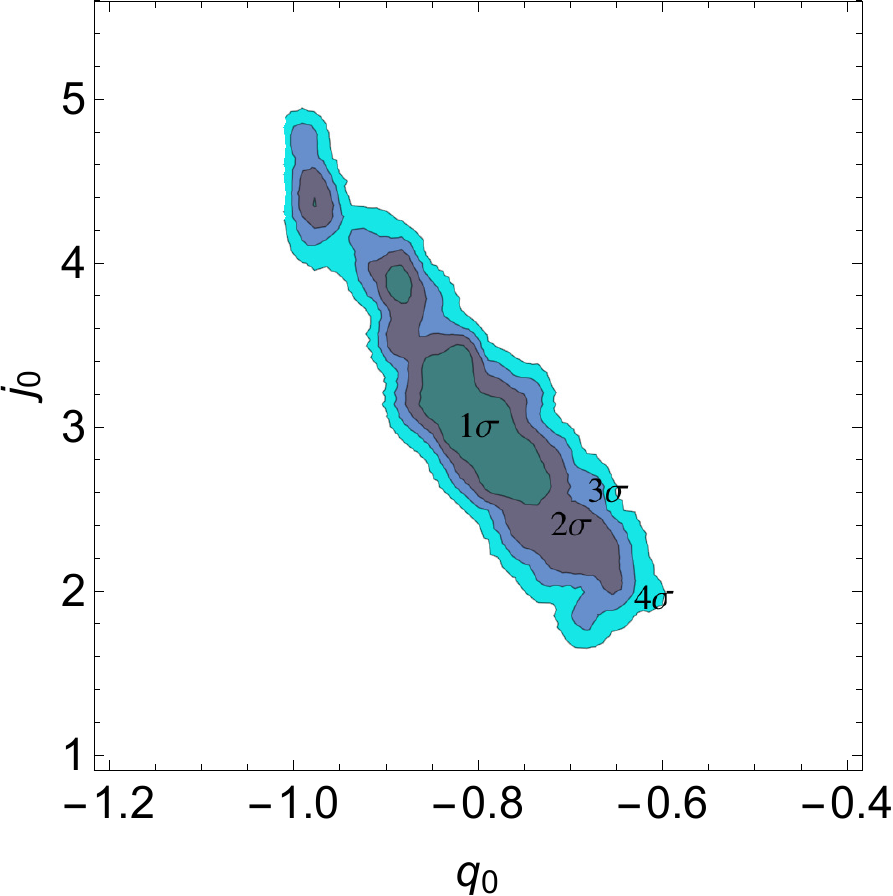} 
\caption{Confidence regions in the ($q_0$-$j_0$) plane as provided by the standard cosmography. The value $j_0=1$ (i.e. $\Lambda$CDM) is off at $>4\sigma$, thus strengthening the results of RL19 using an independent approach.}
\label{conregq0j0}
\end{figure}

\section{Statistical analysis}

\subsection{Cosmological parameters from the standard cosmographic approach}
\label{stdapproach}
To provide statistical constraints on the cosmological parameters, we followed the same procedure as outlined in 
 \citet{demianski12,demianski17b}.
Specifically, we used the  MCMC method and ran three parallel chains. We tested the convergence through the Gelman-Rubin diagnostic approach, which uses as a test probe the reduction factor $R$, i.e.~the square root of the ratio of the variance between-chain and the variance within-chain.  
A large $R$ indicates that the former is substantially greater than the
latter, meaning that a longer simulation is needed. We required that
$R$ converge 
to 1 for each parameter. We set $R - 1$ of the order of $0.05$,
which is more restrictive than the often used and recommended value $R - 1 < 0.1$
for standard cosmological investigations. Moreover, since methods like the MCMC are based on an algorithm that moves randomly in the parameter space, we a priori imposed some basic consistency controls
in the code to reduce the uncertainties
of the cosmographic parameters, requiring that all of the
(numerically) evaluated values of $d_{L}(z)$  be positive. 
We discarded the first $30\%$ of the point iterations at
the beginning of each MCMC run, and thinned the chains that were run many times.
We finally extracted the constraints on cosmographic parameters by coadding the
thinned chains and inferred the median values and the confidence ranges from the marginalisation of the posterior probability distributions of the parameters. 

The Hubble diagram resulting from the combination of SNe Ia (cyan points), quasars (yellow points), and GRBs (black points) is presented in Figure~\ref{hubblediag}. 
The dashed magenta line corresponds to a flat $\Lambda$CDM model fit with $\Omega_{\rm M}= 0.3$, whilst the green dotted line represents the regression fit (equation~\ref{serielum1}) with the inclusion of GRBs and the most updated SNe Ia sample.

A summary of the results is provided in Table~\ref{tabgrb}, whilst in Figure \ref{conregq0j0} we present the resulting confidence regions for $q_0$ and $j_0$.
Our statistical MCMC analysis shows that the jerk parameter, $j_0$, is significantly different from its $\Lambda$CDM value $j_0=1$ at more than 4$\sigma$.

\subsection{Our new cosmographic approach}
\label{ourcosmo}
Figure~\ref{hubblediag} shows the results obtained with the new regression fit (equation~\ref{dl}) considering quasars, GRBs and the {\it Pantheon} sample. A discrepancy between the $\Lambda$CDM model of low-redshift data and the model-independent cosmographic MCMC regression curve obtained with the full data set is observed, in agreement with the findings presented in section~\ref{stdapproach}.  

For comparison, we also present with the black solid line the best MCMC regression curve of the third-order expansion in $\log(1+z)$ published by RL19 and obtained by fitting the approximately $1600$ quasars and the SNe Ia from the JLA survey \citep{betoule2014}. Our new analysis that considers the updated SNe Ia sample and the GRBs further supports this tension with increased statistics. 

\begin{figure}
 \centering
  \includegraphics[width=0.8\linewidth,clip]{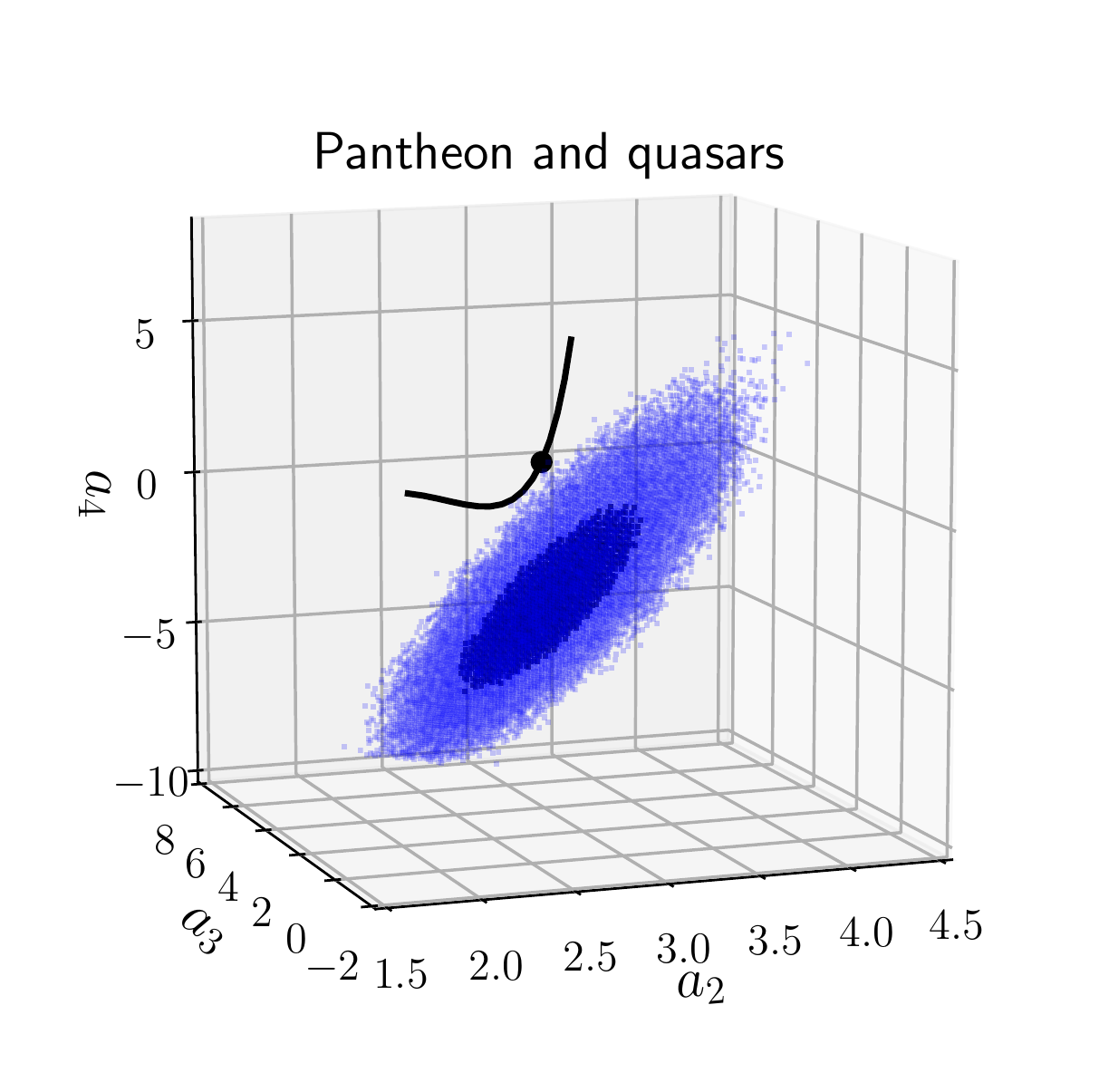}
  \includegraphics[width=0.8\linewidth,clip]{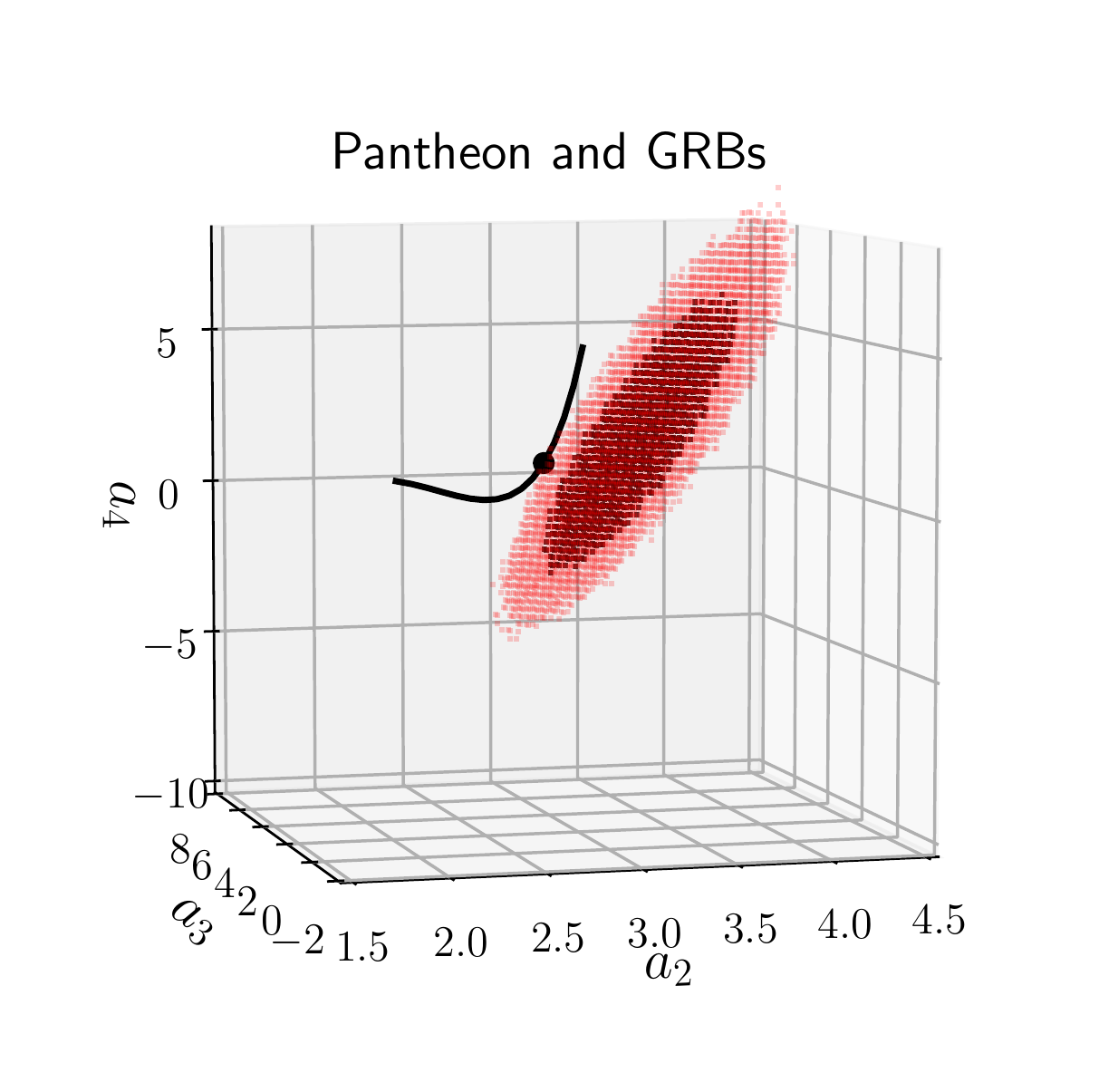}
  \includegraphics[width=0.8\linewidth,clip]{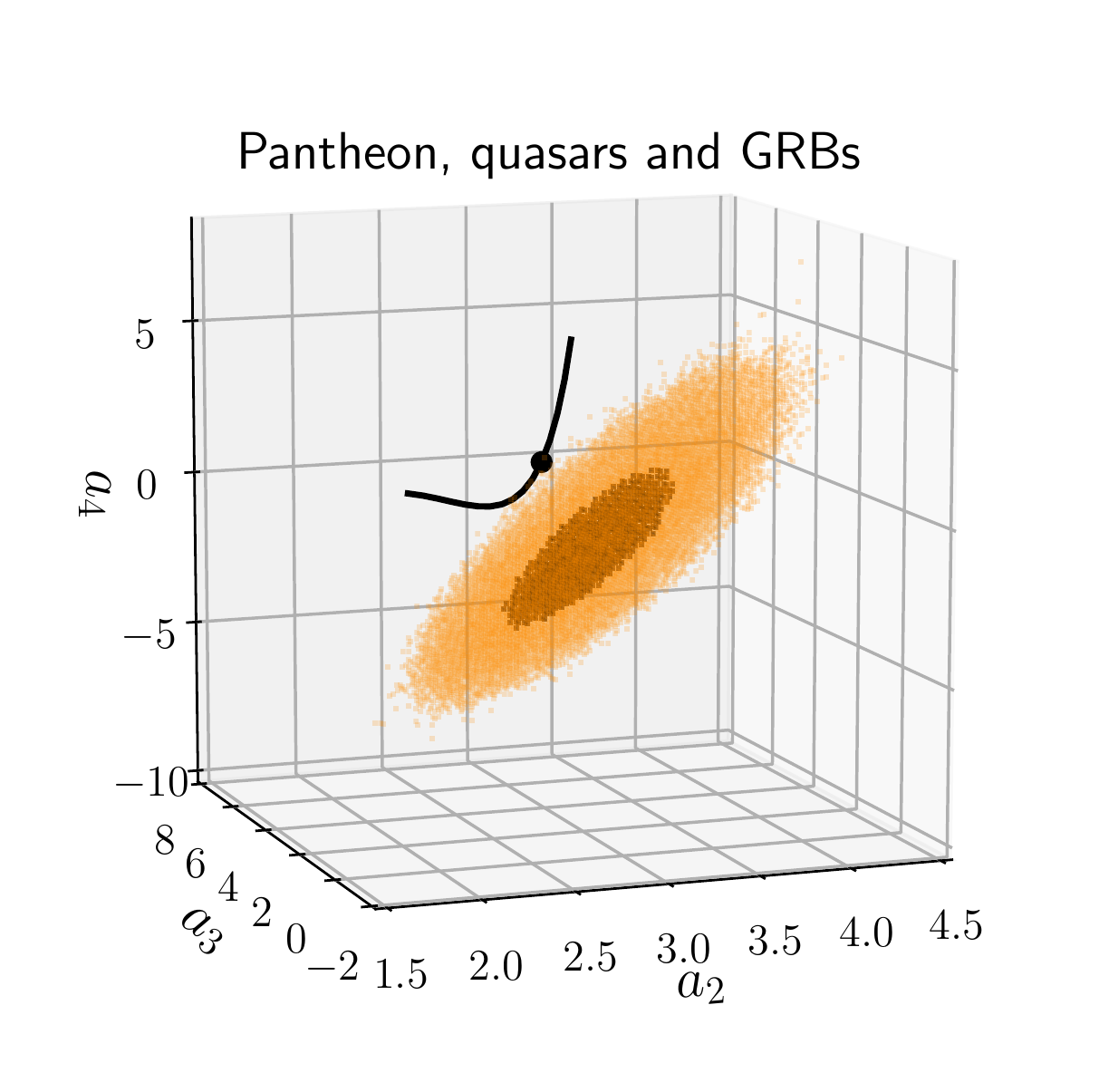}
        \caption{The 1$\sigma$ (dark colours) and 4$\sigma$ (light colours) error contours in the ($a_2, a_3, a_4$) parameter space resulting from fitting the data using equation~\ref{dl} for the samples {\it Pantheon}+quasars and {\it Pantheon}+quasars+GRBs. We plot the 1$\sigma$ and 2$\sigma$ error contours for the {\it Pantheon}+GRBs. The black solid line represents a flat $\Lambda$CDM model with $\Omega_M$ in the interval 0.1--0.9, where the black point marks the value $\Omega_M=0.3$ (i.e. equation~\ref{cosmoparam}). From top to bottom: we present the result from the cosmographic technique described in Sects.~\ref{ourcosmo_theory} and~\ref{ourcosmo}.}
  \label{3d}
\end{figure}

In Figure~\ref{3d} we show the 1$\sigma$ (dark colours) and 4$\sigma$ (light colours) error contours in the ($a_2, a_3, a_4$) parameter space resulting from fitting the data using equation~\ref{dl} for the samples {\it Pantheon}+quasars and {\it Pantheon}+quasars+GRBs, whilst we plot the 1$\sigma$ and 2$\sigma$ error contours for the {\it Pantheon}+GRBs.
The black solid line represents a flat $\Lambda$CDM model with $\Omega_M$ in the interval 0.1--0.9, where the black point marks the value $\Omega_M=0.3$ (i.e. equation~\ref{cosmoparam}). For visualisation purposes, the orientation is chosen to better visualise the discrepancy between the contours and the flat $\Lambda$CDM model. From this figure, it is apparent that the tension of our cosmographic fit with the standard model is at $4\sigma$ for {\it Pantheon}+quasars, and remains significant even with the inclusion of the GRBs (bottom panel of Fig.~\ref{3d}).  
A summary of the results obtained from the MCMC regression analysis of the fourth order is provided in Table~\ref{tab1}.

As these results reflect an average deviation from the flat $\Lambda$CDM over the redshift probed by the different data sets, we can investigate what redshift range drives this tension. 

\begin{figure}
 \centering
  \includegraphics[width=\linewidth,clip]{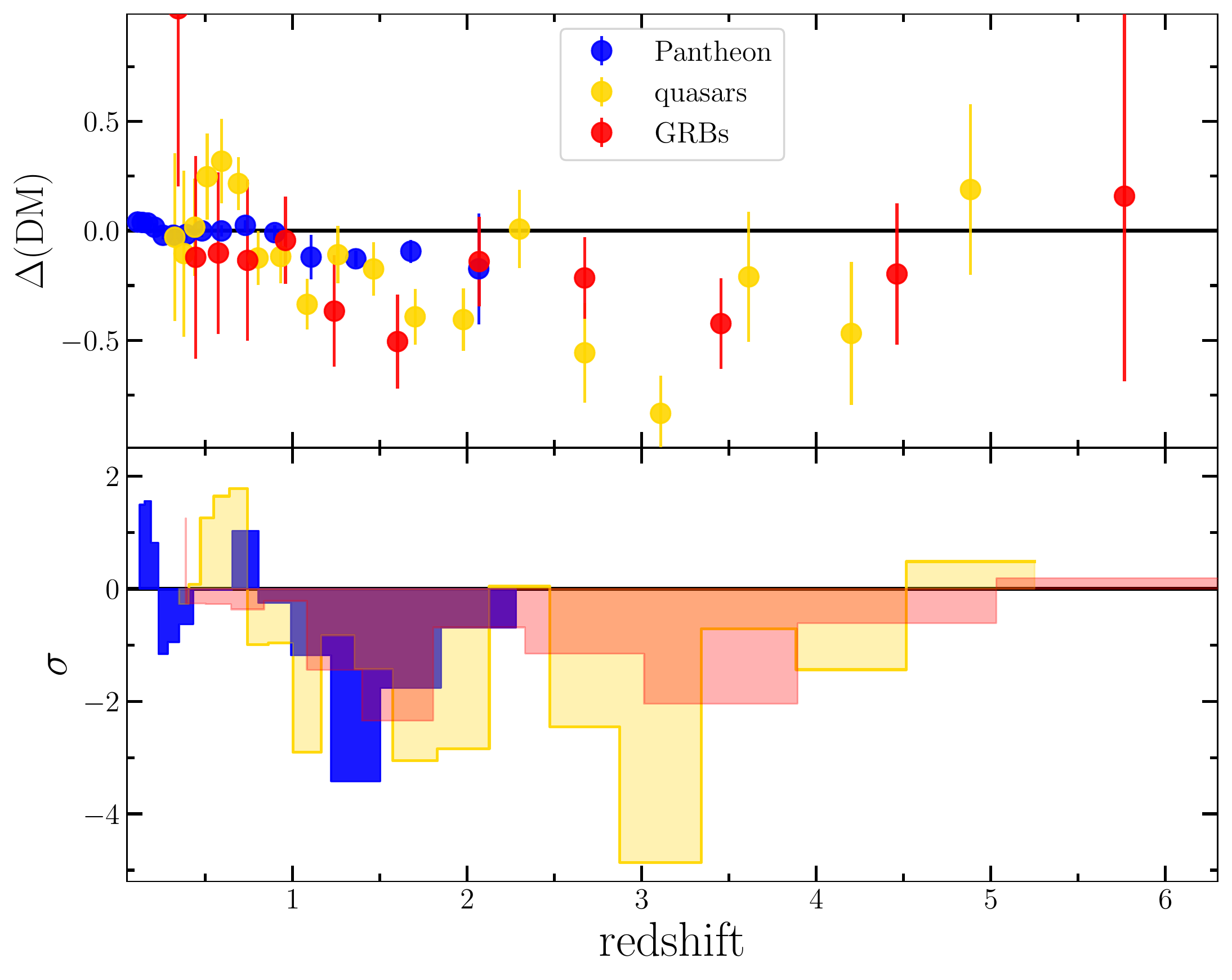}
        \caption{Deviation of the distance modulus from the $\Lambda$CDM with $\Omega_M=0.3$ and normalized to the Pantheon SNe sample up to $z=1$ ($\Delta(\rm DM)=DM-DM_{\rm \Lambda CDM}$) for SNe Ia (blue), quasars (yellow), and GRBs (red). Top panel: Points represent the averages of $\Delta(\rm DM)$ in logarithmic redshift intervals. Bottom panel: Deviations in $\sigma$. We note that the statistical significance of the discrepancy from the $\Lambda$CDM in Fig.~\ref{3d} is not directly comparable to the one plotted here since the MCMC fits in the former are computed over the entire redshift range.}
  \label{sigmas}
\end{figure}

Figure~\ref{sigmas} shows how the deviation from a $\Lambda$CDM (with $\Omega_M=0.3$ and normalized to the {\it Pantheon} SNe sample up to $z=1$) evolves with redshift for the three samples taken singularly. In the top panel we plot the average difference in distance modulus (DM) between the observed DM and the fit assuming a flat $\Lambda$CDM model ($\Delta(\rm DM)=DM-DM_{\rm \Lambda CDM}$) in $\log z$ intervals, whilst the bottom panel represents the same deviation in $\sigma$ units.
Interestingly, all the three samples show that the deviation from the flat $\Lambda$CDM becomes increasingly statistically significant at redshifts higher than 1, with with 23 SNe Ia at $z>1$ at more than $3\sigma$ tension with the flat $\Lambda$CDM model. 

As a final comment, we mention that the tension with the flat $\Lambda$CDM has a peak at $z\simeq3$ (where we have excellent pointed observations from our XMM-Newton large program), but it seems to decrease at $z > 3$ where the data quality reduces. Pointed observations of quasars at $z\simeq4$ are required to verify this trend.


\begin{table}
\caption{\label{tab1}Cosmographic results using the logarithmic polynomial function.}
\centering
\begin{tabular}{lccc}
\hline\hline
sample & $a_2$ & $a_3$ & $a_4$  \\
\hline
\rule{0pt}{3ex}
SNe, quasars, GRBs & $3.205^{+0.165}_{-0.162}$ & $3.564^{+0.916}_{-0.938}$ & $-2.510^{+1.595}_{-1.536}$ \\
\rule{0pt}{3ex}
SNe, quasars & $3.075^{+0.172}_{-0.169}$ & $4.466^{+1.013}_{-1.040}$ & $-3.716^{+1.922}_{-1.852}$ \\
\rule{0pt}{3ex}
SNe, GRBs & $3.304^{+0.186}_{-0.183}$ & $2.069^{+1.1217}_{-1.252}$ & $2.571^{+2.631}_{-2.506}$   \\[0.2cm]
\hline
\end{tabular}
\end{table}

\section{Conclusions}
The spatially flat $\Lambda$CDM cosmology is still the most commonly assumed representation of our Universe within the scientific community, yet there is mounting evidence that the standard $\Lambda$CDM model (or its simplest extension, e.g. $w$CDM) is in tension at more than $4\sigma$ with local direct measurements of the Hubble constant (e.g. \citealt{riess2019} and references therein).

Recently, RL19 also reported on a $\sim$4\,$\sigma$ tension with the $\Lambda$CDM model through a model-independent parametrization of a Hubble Diagram of SNe Ia and quasars. Since this tension could be subject to some level of systematic error, it is paramount to test the $\Lambda$CDM model by making use of model-independent approaches and independent samples over a wide redshift range.

To this aim, we present a joint statistical analysis of the Hubble diagram from SNe Ia, quasars, and GRBs, thus using three completely independent samples of standardisable candles spanning a wide redshift interval, from local objects up to $z\simeq6.7$. 
The statistical analysis was performed in two ways: 1. assuming a traditional cosmographic approach and 2. through a cosmographic fourth-order polynomial fit in $\log(1+z)$ of the Hubble diagram, following RL19. 

From the standard approach, we found that $j_0$ is significantly different from its $\Lambda$CDM value, $j_0=1$, at more than 4$\sigma$.

Pursuing the cosmographic approach, we again confirmed the tension between the best-fit cosmographic parameters and the $\Lambda$CDM model with the joint SNe Ia, quasar, and GRB data set at more than the $4\sigma$ statistical level. We also performed the same analysis by considering the SNe Ia+quasar and the SNe Ia+GRB samples separately, finding that the tension is still statistically significant at the $\sim$$4\sigma$ and $\sim$$2\sigma$ level, respectively. Moreover, we also confirm that this tension becomes statistically significant (above $3\sigma$) only at high redshifts ($z>1$) for SNe Ia and quasars taken independently, whilst this is at $\sim2\sigma$ for GRBs alone.

Summarising, these two complementary (and independent) statistical analyses both confirm a tension with the flat $\Lambda$CDM model at 4$\sigma$ emerging at high redshifts.
Moreover, as the completely independent high-redshift Hubble diagrams of quasars and GRBs are fully consistent with each other, this strongly suggests that the deviation from the standard model is not due to unknown systematic effects but to new physics. Possible extensions of the standard cosmological model will be explored in future works.

\begin{acknowledgements}
We acknowledge financial contribution from the agreement ASI-INAF n.2017-14-H.O. 
SB is supported by NASA through the Chandra award no. AR7-18013X (NAS8-03060) and by the grant HST-AR-13240.009. 
EN acknowledges funding from the EU Horizon 2020 Marie Sk\l{}odowska-Curie grant no. 664931.
\end{acknowledgements}

%
   \bibliographystyle{aa} 
   \bibliography{bibl} 
%

\end{document}